\documentclass[sigconf,10pt,nonacm]{acmart}
\AtBeginDocument{%
  }

\setcopyright{none}
\renewcommand\footnotetextcopyrightpermission[1]{}

\usepackage{amsmath,amsthm}
\usepackage{array}
\usepackage{booktabs}
\usepackage{enumitem}
\usepackage{graphicx}
\usepackage{multirow}
\usepackage{tabularx}
\usepackage{xcolor}
\usepackage[normalem]{ulem}
\usepackage{xurl}
\newif\ifannotated

\newcolumntype{C}[1]{>{\centering\arraybackslash}p{#1}}

\newcommand{\uptwo}{\textcolor{green!55!black}{\(\uparrow\uparrow\)}}
\newcommand{\downone}{\textcolor{red!75!black}{\(\downarrow\)}}
\newcommand{\downtwo}{\textcolor{red!75!black}{\(\downarrow\downarrow\)}}

\newcommand{\resultcell}[2]{%
  \makebox[2.4em][r]{#1}%
  \makebox[1.35em][l]{\,#2}%
}

\annotatedfalse  % Set to annotated mode (use \annotatedtrue for annotated mode \annotatedfalse for clean mode)

\begin{document}

\title[Evidence-Grounded Mapping of Multimodal Human Sensing \\to
Psychological Transdiagnostic Dimensions]{
Evidence-Grounded Mapping of Multimodal Human Sensing to
Psychological Transdiagnostic Dimensions}
% Benchmarking LLMs for Mapping EMA and Passive Sensing to HiTOP Spectrum Scores}
\author{Xiyun Hu$^\dag$, Xiangyuan Xue$^{\dag, \ddag}$, Yuting Lyu$^{\dag, *}$, Hanya Shao$^\#$, and Jingping Nie$^\dag$}
\affiliation{\institution{$^\dag$University of North Carolina at Chapel Hill, $^{\ddag}$University of Auckland, $^*$Purdue University,
\\$^\#$East River Counseling} \country{\{huxiyun, xyxue, ylyu, jingping\}@unc.edu, hanya@eastrivercounseling.com}}
\renewcommand{\shortauthors}{Hu et al.}

\begin{abstract}
Mobile and wearable sensing enables longitudinal observation of behavior, yet translating these signals into meaningful mental health constructs remains difficult. We introduce a clinician-in-the-loop benchmark for evaluating whether large language models (LLMs) can generate evidence-grounded Brief Hierarchical Taxonomy of Psychopathology (B-HiTOP) item profiles from passive sensing, ecological momentary assessment (EMA), and questionnaire evidence. Using the Generalization of Longitudinal Behavior Modeling (GLOBEM) dataset, we construct 14,592 participant-day instances and align multimodal evidence to 29 B-HiTOP items across five spectra. Since GLOBEM lacks B-HiTOP responses, we evaluate evidence compatibility (\texttt{\textbf{C}}) rather than diagnostic accuracy, separating substantive predictions from abstentions when evidence is insufficient for item-level scoring. Two-stage prediction improves \texttt{\textbf{C}} for EMA and questionnaire evidence, but reduces \texttt{\textbf{C}} under passive sensing and combined evidence and produces more conservative score distributions across models, spectra, and evidence settings. Overall, semantic abstraction helps organize heterogeneous self-report evidence while becoming an information bottleneck for indirect behavioral sensing signals.

\end{abstract}




\maketitle
\section{Introduction}

Mobile phones and wearable devices enable longitudinal observation of sleep, mobility, physical activity, social interaction, affect, and momentary self-reports. Prior work in digital phenotyping has shown that passive sensing can characterize behavioral indicators of depressive symptoms and support mental health prediction in naturalistic settings~\citep{saeb2015mobile,chikersal2021detecting}. Most existing studies, however, are formulated around a single disorder, questionnaire-based score, or clinical outcome. These formulations support targeted prediction but provide a limited account of the overlapping and multidimensional organization of psychopathology.

\begin{figure}[t!]
    \centering
    \includegraphics[
        width=\columnwidth,
        keepaspectratio
    ]{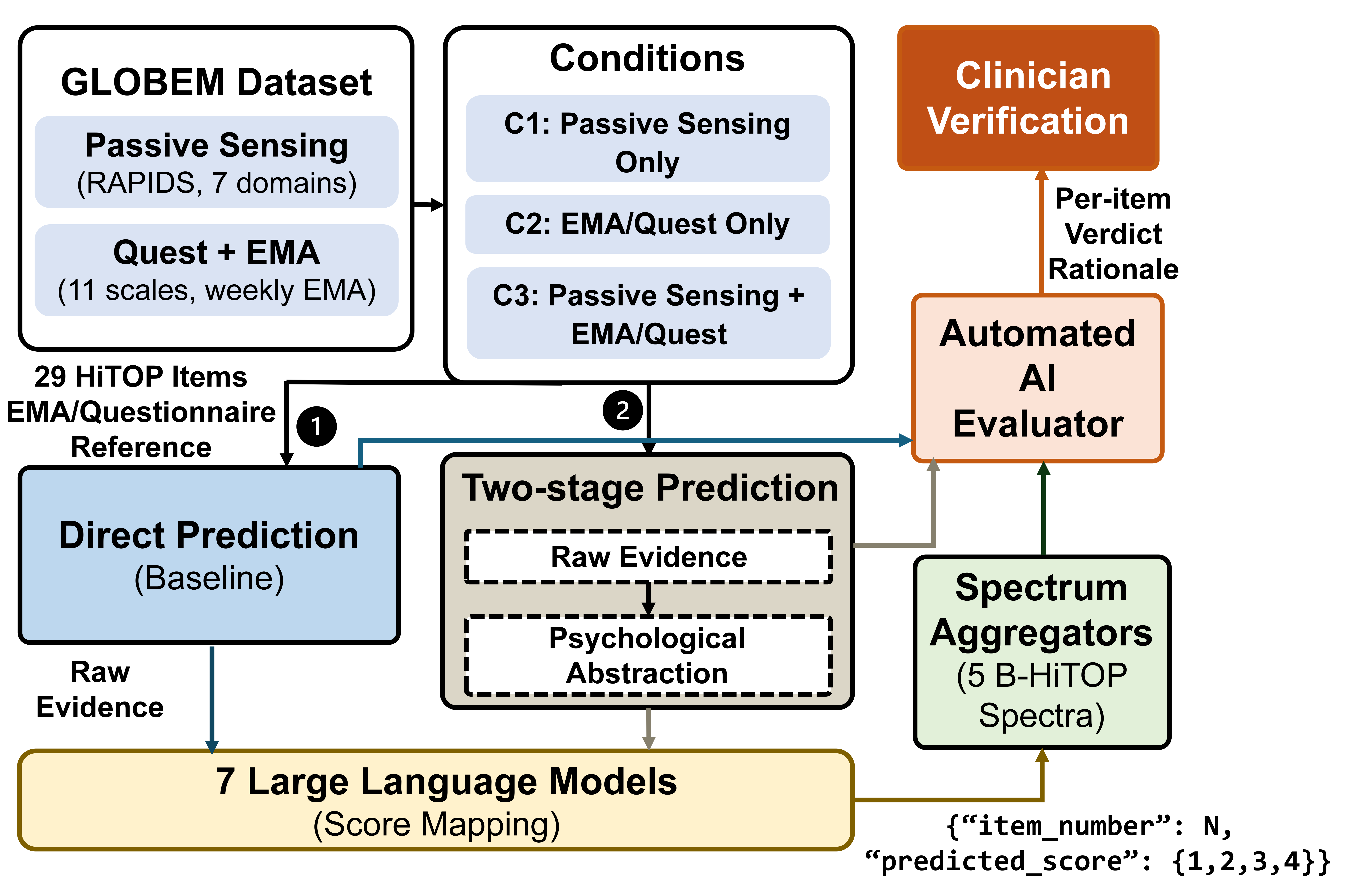}
    \vspace{-0.1in}
    \caption{Overview of the benchmark framework, including three evidence conditions, two prediction pipelines, AI evaluation, and clinician verification.}
    \Description{A benchmark pipeline in which passive sensing, EMA and questionnaire evidence, and combined evidence are processed by direct and two-stage prediction pipelines. The resulting B-HiTOP item scores are assessed by an AI evaluator and verified on a stratified sample by clinicians.}
    \label{fig:architecture}
    \vspace{-0.2in}
\end{figure}

The Hierarchical Taxonomy of Psychopathology (HiTOP) provides a dimensional and transdiagnostic framework that organizes psychopathology into empirically derived spectra rather than discrete diagnostic categories \citep{kotov2017hitop}. A transdiagnostic framework intuitively aligns with human sensing as it accounts for the reality that behavioral observations are rarely specific to a single clinical condition. For instance, behavioral indicators in sensing such as disrupted sleep, reduced mobility, or lower social interaction may relate to internalizing symptoms, somatic symptoms, detachment, or behavioral dysregulation, depending on the surrounding evidence. The Brief Hierarchical Taxonomy of Psychopathology (B-HiTOP) provides a compact item-based measure of broad psychopathology spectra \citep{hegarty2025bhitop}. We therefore use B-HiTOP as a structured target for clinically verifiable transdiagnostic symptom hypotheses.

We use the Generalization of Longitudinal Behavior Modeling (GLOBEM) dataset because it combines multi-year passive sensing, ecological momentary assessment (EMA), and questionnaire measures across repeated study waves~\citep{xu2022globem,xu2023globem}. Compared with StudentLife, which focuses on an intensively observed student cohort, and RADAR-MDD, which is organized around major depressive disorder, GLOBEM provides the broader repeated-wave evidence needed for transdiagnostic comparison~\citep{wang2014studentlife,matcham2019radarmdd}. Its features cover sleep, screen use, physical activity, mobility, proximity, and communication, allowing sensing and self-report evidence to be examined separately and jointly. GLOBEM does not contain observed B-HiTOP responses; each generated score is interpreted as an evidence-grounded hypothesis rather than a reconstruction of an unobserved clinical response. Although the GLOBEM EMA protocol included items that could be mapped to HiTOP constructs, the released dataset provides only aggregate scores, such as a PHQ-4 composite rather than item-level responses. This limits the precision of item-level alignment and motivates our evaluation of evidence compatibility against participant-day evidence instead of diagnostic accuracy. Access to item-level EMA data would improve the specificity of mapping onto the B-HiTOP framework.

Large language models can integrate sensing summaries, self-report measures, and construct definitions, but their use in mental health requires clinically grounded evaluation and alignment~\citep{singhal2023medpalm, hager2024clinical,xia2025convergence}. We construct a clinician-verified benchmark of 14,592 GLOBEM participant-day instances and 29 B-HiTOP items across five spectra. We compare a direct prediction with a two-stage pipeline that first generates a psychological abstraction and then scores items from that representation~\citep{fan2026timesrl}. Both pipelines use the same instances, models, items, response scale, output schema, and evaluation procedure across three evidence settings. A fixed AI evaluator assesses score reasonableness against the original evidence, and licensed clinicians verify a stratified sample of judgement. The study asks two research questions:

\begin{itemize} [leftmargin=1em, itemsep=1pt, topsep=1pt]
    \setlength\itemsep{0.05em}
    \setlength\parskip{0pt}
    \setlength\parsep{0pt}
    \setlength\topsep{0pt}

    \item \textbf{RQ1:} How can evidence compatibility be evaluated in a clinically verifiable way when observed B-HiTOP responses are unavailable?

    \item \textbf{RQ2:} How do direct and two-stage predictions differ across evidence settings, models, spectra, and score distributions?
\end{itemize}

\noindent This paper makes two contributions:
\begin{itemize} [leftmargin=1em, itemsep=1pt, topsep=1pt]
    \setlength\itemsep{0.05em}
    \setlength\parskip{0pt}
    \setlength\parsep{0pt}
    \setlength\topsep{0pt}

    \item Based on GLOBEM~\cite{xu2023globem}, we construct 14,592 participant-day instances and align GLOBEM evidence to 29 clinician-verified B-HiTOP items across five spectra.
    
    \item We provide a controlled comparison of (1) \emph{direct} and (2) \emph{two-stage} prediction, showing semantic abstraction improves compatibility for EMA and questionnaire evidence, but reduces compatibility when passive sensing evidence is used and produces more conservative score distributions.
\end{itemize}

\section{Benchmark Construction}

\textbf{Task and participant-day construction.}
We formulate the benchmark as an evidence-grounded transdiagnostic profiling task. For each participant-day, a model receives passive sensing evidence, EMA and questionnaire evidence, or both, and scores each retained B-HiTOP item from 1 (``not at all'') to 4 (``a lot''). Scores are treated as evidence-grounded symptom hypotheses rather than diagnoses or reconstructions of unobserved responses; the benchmark evaluates reasonableness given evidence rather than diagnostic accuracy. Instances are constructed by matching daily feature records with same-week EMA observations and pre-survey data. Thus, the prediction unit is a participant-day, while EMA provides week-level self-report context. All evidence settings use the same 14,592 instances (Wave 1: 3879; Wave 2: 4012; Wave 3: 2547; Wave 4: 4154), ensuring that comparisons vary the model input rather than the sample. Table~\ref{tab:data_selection} summarizes the data selection and wave-level instance counts.

\noindent\textbf{Evidence representation.}
Passive sensing evidence comprises RAPIDS features from sleep, screen use, physical activity, mobility, Bluetooth proximity, call behavior, and Wi-Fi context. Features exceeding 80\% missingness are removed within each wave. Each participant-day is summarized in natural language, including feature availability, missingness, wave-level cohort comparisons, and the largest standardized deviations with observed and reference values. Depending on the feature, sensing variables describe daily behavior or historical windows. EMA values are aligned to the week without further aggregation and include affect, PHQ-4 symptoms, perceived stress, and anxiety when available. Questionnaire evidence adds wave-level context from 11 measures covering depression, anxiety, stress, somatic symptoms, loneliness, social support, emotion regulation, resilience, alcohol-related consequences, coping, and mindfulness. A reference sheet preserves source timing, score intervals, scoring direction, and wave-specific instrument variations.

\begin{table}[t!]
\centering
\caption{GLOBEM record retention by study wave.}
\label{tab:data_selection}

\setlength{\tabcolsep}{6pt}
\renewcommand{\arraystretch}{1.08}

\begin{tabular}{lrr}
\toprule
Wave
& Retained / RAPIDS days
& Retention \\
\midrule
Wave 1 & 3,879 / 14,260 & 27.2\% \\
Wave 2 & 4,012 / 21,146 & 19.0\% \\
Wave 3 & 2,547 / 14,111 & 18.0\% \\
Wave 4 & 4,154 / 20,085 & 20.7\% \\
\midrule
\textbf{Total}
& \textbf{14,592 / 69,602}
& \textbf{21.0\%} \\
\bottomrule
\end{tabular}
\end{table}

\noindent\textbf{Clinician-verified evidence alignment.} Since GLOBEM was not designed as a B-HiTOP administration study, we construct an item-level alignment between the available evidence and B-HiTOP constructs. An item is retained only when sensing, EMA, or questionnaire measures provide a clinically defensible basis for reasoning about it. Direct support indicates a closely matched measure, whereas partial support indicates related but less specific evidence. The benchmark retains 29 of the original 45 items across five spectra, including 8 directly supported and 21 partially supported items. Antagonism is excluded since GLOBEM provides insufficient item-level evidence for that specific spectrum.

\noindent\textbf{Prediction pipelines.}
We compare a direct prediction baseline with a two-stage pipeline under controlled conditions. Both use the same participant-day evidence, predictor model, retained items, response scale, output schema, and evaluation procedure. In the direct baseline, the model assigns scores to all 29 retained items from the complete evidence block in a single call. In the two-stage pipeline, the same evidence is first converted into a structured psychological abstraction covering the five B-HiTOP spectra and an overall summary. The model is instructed to rely only on the supplied evidence and to indicate when a domain cannot be assessed. The same model then assigns all item scores using only this abstraction, without access to the original evidence. Both pipelines use the same item wording and scoring format, allowing the intermediate representation to be evaluated directly.

\section{Model Inference and Evaluation Design}

\noindent\textbf{Evidence settings.}
For each participant-day, we construct three evidence conditions. The first setting contains passive sensing evidence only, including behavioral features derived from phone and wearable data. The second contains EMA and questionnaire evidence only, capturing self-report symptoms, affective states, and broader wave-level psychological context. The third combines both evidence blocks. All settings use the same task instruction, B-HiTOP item wording, response scale and reference, and output schema for each instance. Only the evidence block varies, and both prediction pipelines are evaluated under all three conditions. This design enables us to isolate the impact of evidence modality while keeping the participant-day, scoring target, and model instructions fixed. It also tests whether semantic abstraction operates differently when the input evidence is self-reported, behaviorally sensed, or multimodal.  

\noindent\textbf{Predictor models and output format.}
We evaluate seven LLMs: DeepSeek V4 Pro, Grok 4.3, Kimi K2.6, Llama 4, Mistral Large 3, GPT 5.5, and Lingshu 32B ~\citep{deepseek2026v4,xai2026grok43,moonshot2026kimi26,
meta2025llama4,mistral2025large3,openai2026gpt55,lingshu2025}. The direct and two-stage pipelines use pipeline-specific prompts but share the same B-HiTOP item wording, response scale, diagnostic safety instruction, and scoring output schema. In the direct pipeline, all 29-item scores are generated from the original evidence in a single inference call. In the two-stage pipeline, Stage 1 generates a structured psychological abstraction, and Stage 2 generates all 29-item scores in a separate call using the Stage 1 abstraction. Both item-scoring calls use the following strict JSON format:

\begin{center}
\resizebox{0.96\columnwidth}{!}{%
\texttt{\{"item\_number": N, "predicted\_score": \{1, 2, 3, 4\}\}}%
}
\end{center}

\begin{figure}[t!]
    \centering
    \includegraphics[
        width=\columnwidth,
        keepaspectratio
    ]{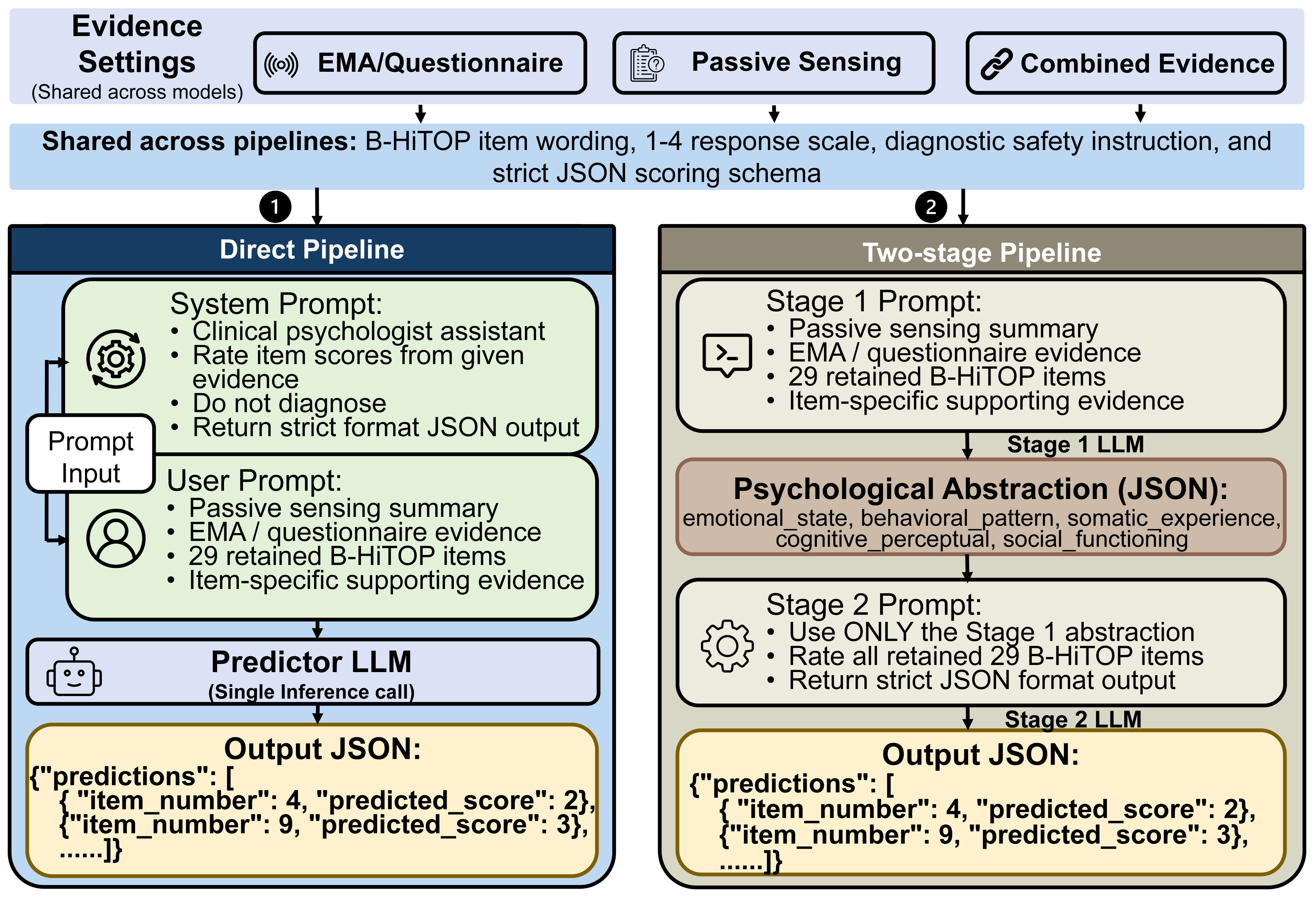}
    \vspace{-0.1in}
    \caption{Prompt and inference design for the (1) \emph{direct} and (2) \emph{two-stage} B-HiTOP prediction pipelines.}
    \Description{The direct pipeline maps the original participant-day evidence directly to 29 B-HiTOP scores. The two-stage pipeline first produces a structured psychological abstraction and then maps that abstraction to the same 29 scores using the same output schema.}
    \label{fig:prompt_design}
    \vspace{-0.2in}
\end{figure}

Scoring outputs are marked as invalid if they cannot be parsed, omit items, duplicate items, or contain out of range scores. Invalid outputs are recorded as format failures and are not manually corrected. For spectrum-level analysis, item scores are averaged within their corresponding B-HiTOP spectra. The primary evaluation remains item-level because spectrum averages can hide construct mismatch and uneven evidence coverage. Figure~\ref{fig:prompt_design} summarizes the prompt inputs, intermediate
representation, and output format used by the direct and two-stage prediction pipelines.

\noindent\textbf{Evidence-grounded evaluation.} Since GLOBEM does not contain observed B-HiTOP responses, we evaluate predictions against the available evidence rather than a ground truth score. For both pipelines, we use a fixed GPT 5.4 AI evaluator~\citep{openai2026gpt54} that receives participant-day evidence, the B-HiTOP item wording, reference sheet, and the predicted score. The evaluator does not receive the abstraction, allowing both pipelines to be assessed against the same evidence.

The evaluator assesses whether each predicted score is consistent with the evidence and clinically plausible for the item. Its judgment and rationale are post-processed using the evaluator judgement rubrics in Table~\ref{tab:evaluation_rubric}. A prediction is classified as \texttt{reasonable} when evidence supports its direction and severity. It is classified as \texttt{unreasonable} when it contradicts relevant evidence or assigns a score from 2 to 4 without evidence. When no relevant evidence is available and the model assigns score 1, the prediction is classified as an \textit{abstention} rather than as evidence of symptom absence. This distinction is necessary because a B-HiTOP score of 1 means ``not at all'', whereas missing or insufficient evidence does not establish that the symptom is absent.

\begin{table}[t!]
\centering
\caption{Evaluator judgement post-processing rubrics.}
\label{tab:evaluation_rubric}
\vspace{-0.1in}
\footnotesize
\setlength{\tabcolsep}{4pt}
\renewcommand{\arraystretch}{1.12}

\begin{tabularx}{0.95\columnwidth}{
    @{}
    >{\raggedright\arraybackslash}X
    >{\centering\arraybackslash}p{1.15cm}
    >{\raggedright\arraybackslash}p{1.65cm}
    @{}
}
\toprule
Evidence condition
& Score
& Outcome \\
\midrule

Relevant evidence and clinically plausible
& Any
& \texttt{Reasonable} \\

Relevant evidence but clinically implausible or contradictory
& Any
& \texttt{Unreasonable} \\

No relevant evidence
& 2--4
& \texttt{Unreasonable} \\

No relevant evidence
& 1
& \texttt{Abstention} \\

\bottomrule
\end{tabularx}
\end{table}

For analysis group \(g\), let \(R_g\), \(U_g\), and \(A_g\) denote the numbers of reasonable, unreasonable, and abstained predictions, respectively. We define evidence compatibility as

\[
C_g = \frac{R_g}{R_g + U_g}.
\]

Abstentions are excluded because they reflect insufficient evidence for substantive item scoring rather than a clinically supported low severity prediction. Thus,  \texttt{\textbf{\(C_g\)}} measures compatibility among predictions for which the available evidence was judged sufficient to support or contradict the assigned score. We interpret  \texttt{\textbf{\(C_g\)}} together with predicted score distributions to effectively distinguish evidence compatibility from prediction coverage and severity differentiation.

\noindent\textbf{Clinician verification.} Following the AI-based evaluation, three licensed clinicians verify the evaluator judgments on a stratified sample of 300 evaluation instances. The sample is stratified across pipelines, predictor models, evidence settings, waves, spectra, and evaluator outcomes. Each instance includes the participant-day evidence, one model's  29-item predictions, evaluator judgments, and evaluator rationales. The clinicians assess whether each evaluator judgment is clinically warranted. This verification does not create participant symptom labels. Instead, it evaluates whether the LLM-based reasonableness assessment is clinically defensible. The protocol balances scalability with clinical accountability, which is necessary for evaluating mental health profiling systems without observed B-HiTOP ground truth.

\section{Results}
\noindent\textbf{Prediction Pipeline Effects Across Evidence Settings.} At the pipeline level, we compare direct and two-stage prediction across the three evidence settings. Evidence compatibility  \texttt{\textbf{C}}  was computed after excluding abstentions, reflecting compatibility among substantive predictions rather than conservative minimum scores. As shown in the average row of Table~\ref{tab:model_level_results}, the two-stage pipeline increased mean  \texttt{\textbf{C}}  from 7.6\% to 53.4\% under EMA and questionnaire evidence. In contrast,  \texttt{\textbf{C}}  decreased from 80.6\% to 21.8\% under passive sensing and from 77.8\% to 58.2\% under combined evidence.

This pattern suggests that intermediate abstraction was most useful for heterogeneous self-report evidence. In this setting, abstraction likely organized distributed affect, stress, anxiety, and questionnaire signals into a representation better suited for item-level scoring. The same benefit did not extend to passive sensing. When behavioral features were the main evidence source, converting sensing summaries into psychological abstractions reduced compatibility, suggesting that feature magnitudes, missingness patterns, and weak behavioral deviations may be important for scoring. The two-stage pipeline should therefore be interpreted as a modality-dependent representation choice rather than a universally superior prediction strategy.

\begin{table}[t!]
\centering
\caption{Model-level evidence compatibility ( \texttt{\textbf{C}} ) across pipelines and evidence settings, excluding abstentions.}
\label{tab:model_level_results}

\scriptsize
\setlength{\tabcolsep}{1.5pt}
\renewcommand{\arraystretch}{1.07}

\resizebox{\columnwidth}{!}{%
\begin{tabular}{
    l@{\hspace{1.8em}}
    l
    C{1.48cm}
    C{1.00cm}
    C{1.00cm}
}
\toprule
Model
& Pipeline
& \shortstack{EMA/\\Questionnaire}
& Sensing
& Combined \\
\midrule

\multicolumn{5}{l}{\textit{Proprietary Models}} \\
\addlinespace[1pt]

\multirow{2}{*}{DeepSeek V4 Pro}
& Direct
& \resultcell{6.9}{}
& \resultcell{70.4}{}
& \resultcell{69.8}{} \\
& two-stage
& \resultcell{36.8}{\uptwo}
& \resultcell{6.5}{\downtwo}
& \resultcell{40.9}{\downtwo} \\
\addlinespace[2pt]

\multirow{2}{*}{Grok 4.3}
& Direct
& \resultcell{7.5}{}
& \resultcell{82.9}{}
& \resultcell{81.1}{} \\
& two-stage
& \resultcell{40.5}{\uptwo}
& \resultcell{0.7}{\downtwo}
& \resultcell{46.4}{\downtwo} \\
\addlinespace[2pt]

\multirow{2}{*}{Kimi K2.6}
& Direct
& \resultcell{5.4}{}
& \resultcell{83.9}{}
& \resultcell{77.5}{} \\
& two-stage
& \resultcell{46.4}{\uptwo}
& \resultcell{7.4}{\downtwo}
& \resultcell{54.5}{\downtwo} \\
\addlinespace[2pt]

\multirow{2}{*}{Mistral Large 3}
& Direct
& \resultcell{6.6}{}
& \resultcell{65.9}{}
& \resultcell{79.0}{} \\
& two-stage
& \resultcell{72.9}{\uptwo}
& \resultcell{61.5}{\downone}
& \resultcell{77.2}{\downone} \\
\addlinespace[2pt]

\multirow{2}{*}{GPT 5.5}
& Direct
& \resultcell{7.0}{}
& \resultcell{80.9}{}
& \resultcell{70.9}{} \\
& two-stage
& \resultcell{30.7}{\uptwo}
& \resultcell{1.2}{\downtwo}
& \resultcell{41.8}{\downtwo} \\

\midrule
\multicolumn{5}{l}{\textit{Open-Weight Models}} \\
\addlinespace[1pt]

\multirow{2}{*}{Llama 4}
& Direct
& \resultcell{6.2}{}
& \resultcell{88.7}{}
& \resultcell{81.5}{} \\
& two-stage
& \resultcell{83.1}{\uptwo}
& \resultcell{45.6}{\downtwo}
& \resultcell{81.5}{} \\
\addlinespace[2pt]

\multirow{2}{*}{Lingshu 32B}
& Direct
& \resultcell{13.8}{}
& \resultcell{91.4}{}
& \resultcell{84.7}{} \\
& two-stage
& \resultcell{63.7}{\uptwo}
& \resultcell{30.0}{\downtwo}
& \resultcell{64.8}{\downtwo} \\

\midrule

\multirow{2}{*}{\textbf{Average}}
& \textbf{Direct}
& \resultcell{\textbf{7.6}}{}
& \resultcell{\textbf{80.6}}{}
& \resultcell{\textbf{77.8}}{} \\
& \textbf{two-stage}
& \resultcell{\textbf{53.4}}{\uptwo}
& \resultcell{\textbf{21.8}}{\downtwo}
& \resultcell{\textbf{58.2}}{\downtwo} \\

\bottomrule
\end{tabular}%
}

\vspace{2pt}
\begin{minipage}{\columnwidth}
\scriptsize
\textit{Note.} Green and red arrows indicate increases and decreases from direct to two-stage prediction, respectively. Double arrows indicate changes of at least 10 pp.
\end{minipage}

\end{table}

\begin{figure}[t!]
    \centering
    \includegraphics[
        height = 8cm,
        width=\columnwidth,
        keepaspectratio
    ]{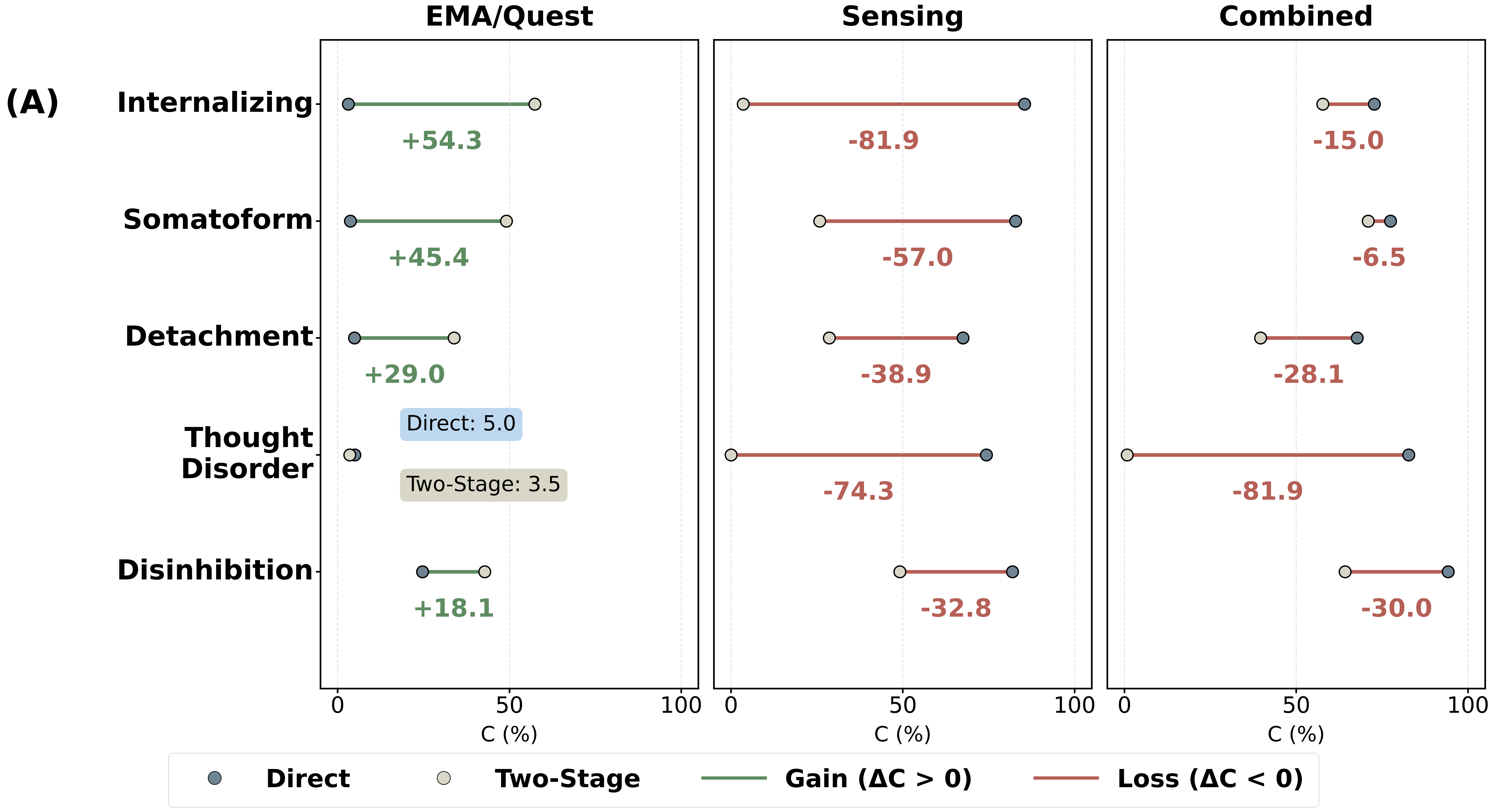}

    \includegraphics[
        width=\columnwidth,
        keepaspectratio
    ]{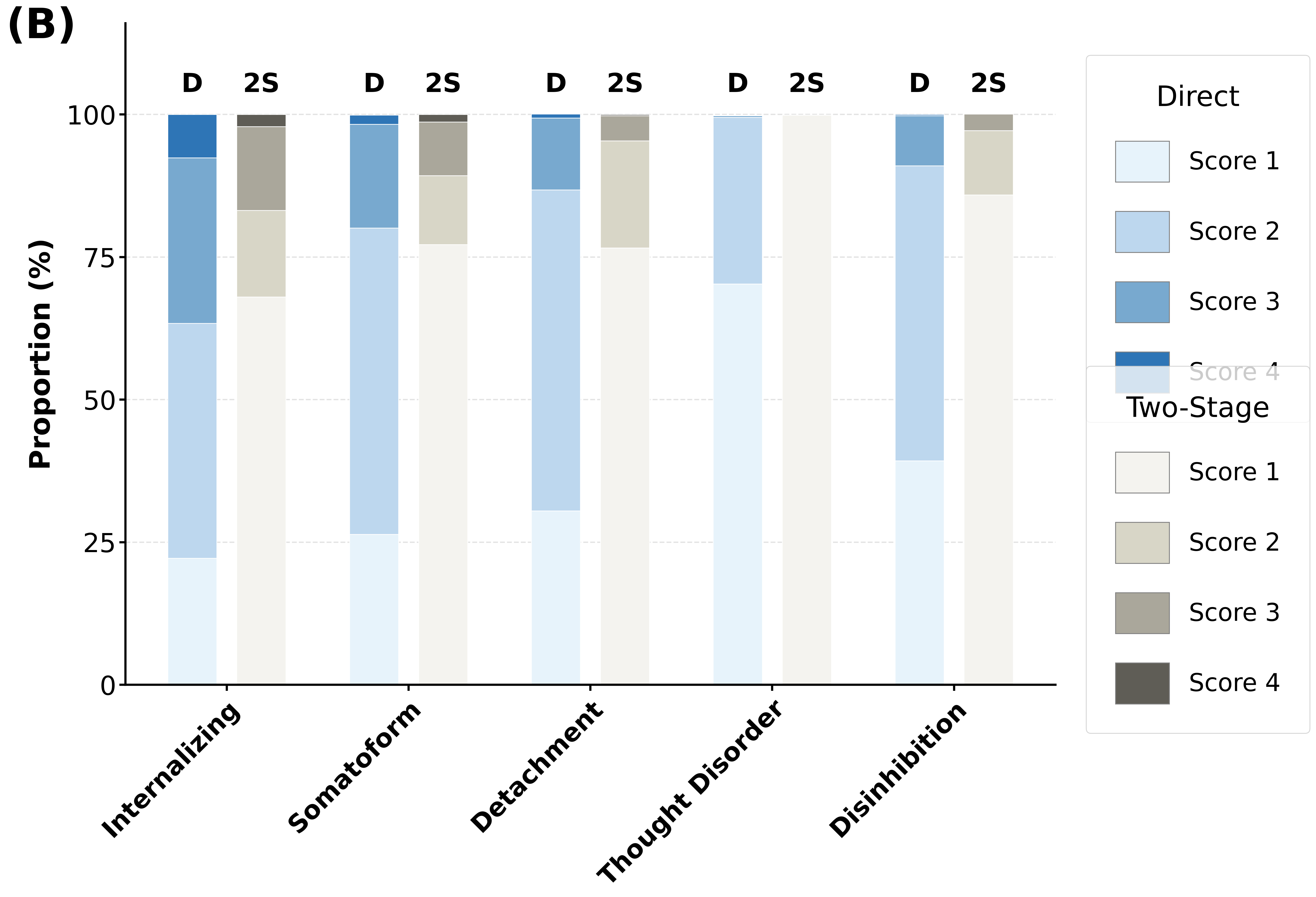}

    \vspace{-0.1in}
    \caption{Spectrum-level effects of semantic abstraction. (A) Changes in evidence compatibility (\(\Delta C\)) by spectrum and condition. (B) Score distributions by spectrum and pipeline.}
    \Description{Panel A compares the change in evidence compatibility from direct to two-stage prediction across B-HiTOP spectra and evidence settings. Panel B compares predicted score distributions across spectra for the direct and two-stage pipelines.}
    \label{fig:spectrum_score_results}
    \vspace{-0.2in}
\end{figure}

\noindent\textbf{Model-Level Effects.} The model-level results in Table~\ref{tab:model_level_results} show that the EMA and questionnaire advantage of the two-stage pipeline was consistent across all seven predictor models. This consistency suggests that the effect was not driven by a single model family or by one particularly strong predictor. Instead, the intermediate representation provided a useful scaffold for translating self-report evidence into B-HiTOP item scores. Evidence compatibility decreased for every predictor model under passive sensing after abstraction, although the magnitude varied across models. Combined evidence also showed a largely negative pattern, with most models losing compatibility under the two-stage pipeline and only limited exceptions. These results suggest that model performance depends jointly on the input modality and the form in which evidence is represented. Semantic abstraction can improve scoring when the evidence is already close to psychological constructs, but it can become a bottleneck when the task requires preserving structured behavioral detail.

\noindent\textbf{Spectrum, Score, and Wave Effects.} Figure~\ref{fig:spectrum_score_results} shows that the effect of semantic abstraction varied across spectra and evidence conditions. Under EMA and questionnaire evidence, the two-stage pipeline improved evidence compatibility for \emph{Internalizing}, \emph{Somatoform}, \emph{Detachment}, and \emph{Disinhibition}, but showed essentially no benefit for \emph{Thought Disorder}, which was weakly represented in GLOBEM. In contrast, semantic abstraction reduced compatibility across all retained spectra under passive sensing and combined evidence, with especially pronounced declines for \emph{internalizing} and \emph{Thought Disorder}. This pattern suggests that abstraction can consolidate fragmented self-report signals into a representation that supports item-level scoring, while simultaneously removing clinically relevant detail from indirect behavioral evidence.

The two-stage pipeline also produced a shift toward score 1 across all spectra, with the strongest floor effect observed for Thought Disorder. Since abstentions are excluded from \(C\), this shift does not mechanically increase evidence compatibility. Instead, it indicates reduced severity differentiation and lower prediction coverage. Evidence compatibility and score distributions should therefore be interpreted jointly. Abstraction may improve the coherence of evidence representation in specific settings while narrowing the range of symptom hypotheses generated by the model.

\begin{figure}[t!]
    \centering
    \includegraphics[
        width=\columnwidth,
        keepaspectratio
    ]{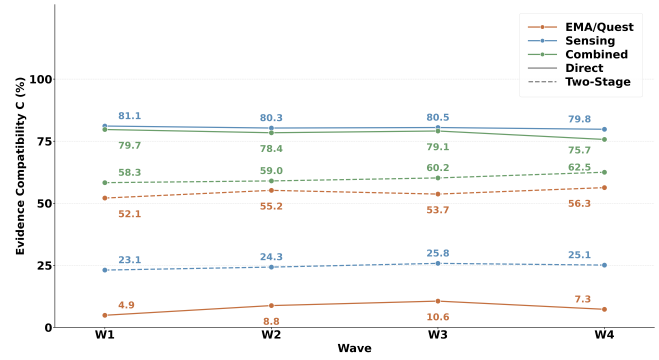}
    \vspace{-0.1in}
    \caption{Evidence compatibility across waves by evidence condition and prediction pipeline.}
    \Description{A line chart comparing evidence compatibility across four GLOBEM waves for direct and two-stage prediction under passive sensing, EMA and questionnaire, and combined evidence conditions.}
    \label{fig:wave_results}
    \vspace{-0.2in}
\end{figure}

The modality-dependent effects were consistent across Waves 1--4, as shown in Figure~\ref{fig:wave_results}. Direct prediction maintained high compatibility under passive sensing and combined evidence but remained weak under EMA and questionnaire evidence. The two-stage pipeline exhibited the opposite pattern, preserving a substantial advantage for self-report evidence while remaining less compatible with passive sensing across waves. Although the wave-level analysis is descriptive, the limited between-wave variation indicates that the main findings are unlikely to be driven by a single collection period and instead reflect stable differences in how the two pipelines represent and use each evidence modality.

\section{Discussion and Conclusion}
\label{sec:conclusion_discussion}

\noindent\textbf{Modality-dependent abstraction.}
Our results indicate that semantic abstraction is beneficial only under specific evidence conditions. Specifically, (i) for EMA and questionnaire evidence, the two-stage pipeline improved compatibility by reorganizing distributed self-report signals into a coherent intermediate representation for item-level scoring; (ii) this benefit did not extend to passive sensing or combined evidence, where abstraction reduced compatibility and may have discarded information contained in feature magnitudes, missingness patterns, weak behavioral deviations, or cross-spectrum sensing associations; and (iii) semantic abstraction should be viewed as a modality-dependent representation choice rather than a uniformly superior prediction strategy.

\noindent\textbf{Limitations and future work.}
This benchmark evaluates evidence compatibility rather than agreement with observed B-HiTOP responses, which are not available in GLOBEM. Abstention-excluded  \texttt{\textbf{C}} measures the reasonableness of substantive predictions but does not capture prediction coverage. Consequently, a pipeline can achieve compatible predictions while frequently defaulting to the minimum score. This limitation is especially relevant for the two-stage pipeline, which produced more concentrated score-1 distributions and near-floor predictions for weakly represented spectra such as Thought Disorder. Construct coverage is also uneven, since only 29 of 45 B-HiTOP items were retained and several are supported by indirect evidence. Future work should jointly report compatibility, abstention behavior, and score distributions; incorporate multiple clinical evaluators; and test whether these findings generalize across datasets, populations, and longitudinal splits.

\noindent\textbf{Conclusion.}
We presented a clinician-in-the-loop benchmark for evidence-grounded B-HiTOP profiling from passive sensing, EMA, and questionnaire evidence. The clinician-in-the-loop design operationalizes evidence compatibility as a clinically verifiable evaluation framework for settings where observed B-HiTOP responses are unavailable. Across 14,592 GLOBEM participant-day instances, seven LLMs, 29 retained items, and three evidence settings, two-stage abstraction improved compatibility for EMA and questionnaire evidence but reduced compatibility for passive sensing and combined evidence. It also produced more conservative score distributions. These findings show that intermediate psychological abstraction can improve self-report organization, but may act as an information bottleneck for indirect behavioral sensing. Future LLM-based mental health profiling systems should match their representation strategy to the modality, specificity, and evidential strength of the available data.

\bibliographystyle{ACM-Reference-Format}
\bibliography{bib/references}

%\appendix
\end{document}